\documentclass{elsart5}

\def\a{\alpha}		 		   	
\def\e{\epsilon}         \def\s{\sigma}       	   	
\def\l{\lambda}	         					
	         \def\W{\Omega}

\def\AJ{A_{\rm J}}       \def\AN{A_{\rm N}}	   
\def\RN{{ R}_{\rm N}}    \def\RF{{ R}_{\rm F}}			
\def\lso{\eta_{\rm so}}       
\def\lN{\l_{\rm N}}	 \def\lF{\l_{\rm F}}	   
\def\sN{\s_{\rm N}}             \def\eF{\e_{\rm F}} 	 
\def\muN{\mu_{\rm N}}    \def\dmuN{\delta\mu_{\rm N}} 
\def\rN{\rho_{\rm N}}    \def\rF{\rho_{\rm F}}     \def\sN{\s_{\rm N}}	
\def\dN{d_{\rm N}}            
            
\def\us{\uparrow}         \def\ds{\downarrow}	
	 \def\kF{k_{\rm F}}	   
\def\[{\left[}           \def\]{\right]}	   
\def\({\left(}           \def\){\right)}	   
\def\<{\langle}          \def\>{\rangle}	   
	          \def\tsf{\tau_{\rm sf}}
\def\timp{\tau_{\rm imp}}\def\Vimp{V_{\rm imp}}

\def\pF{p_{\rm F}}	 \def\PT{P_{\rm T}}        
	          
\def\aH{\alpha_{\rm H}}	   		
\def\aSJ{\aH^{\rm SJ}}
\def\aSS{\aH^{\rm SS}}
\def\nimp{n_{\rm imp}} 
\def\ddr{\nabla_{}} 
\usepackage{graphics}
\usepackage{epsfig}
\usepackage{amssymb}

\begin{document}

\begin{frontmatter}

\title{Nonlocal spin Hall effect and spin-orbit interaction in nonmagnetic metals}
%%\thanks[tit1]{Title footnote}

\author{S. Takahashi\corauthref{cor1}}
\ead{takahasi@imr.tohoku.ac.jp}
%% \ead[url]{home page}
%% \thanks[label1]{author footnote}
\corauth[cor1]{S. Takahashi}
\author{S. Maekawa}
\address{Institute for Materials Research, Tohoku University, Sendai, 980-8577, Japan}
\address{CREST, Japan Science and Technology Agency,  Kawaguchi, 332-0012, Japan}
%% \thanks[label2]{aff footnote}
\received{12 June 2005}
\revised{13 June 2005}
\accepted{14 June 2005}
%use optional labels to link authors explicitly to addresses:

\begin{abstract}
%---------------------------------------------------------------
Spin Hall effect in a nonlocal spin-injection device is theoretically 
studied.
Using a nonlocal spin-injection technique, a pure spin current is 
created in a nonmagnetic metal (N).  The spin current flowing in N is 
deflected by spin-orbit scattering to induce the Hall current in the 
transverse direction and accumulate charge at the edges of N, yielding 
the spin-current induced Hall effect.
We propose a method for extracting the spin-orbit coupling parameter 
in nonmagnetic metals via the nonlocal spin-injection technique.
%---------------------------------------------------------------
\end{abstract}


\end{frontmatter}

%%%%%%%%%%%%%%%%%%%%%%%%%%%%%%%
%\section{Introduction}\label{}
%%%%%%%%%%%%%%%%%%%%%%%%%%%%%%%

There has been growing interest in spin transport in magnetic 
nanostructures, because of potential applications to spin electronic
devices 
\cite{book}.  
Recent experimental studies have demonstrated that 
the spin polarized carriers injected from a ferromagnet (F) into 
a nonmagnetic material (N) such as a normal metal 
\cite{jedema,kimura,garzonPRL,godfrey}
and superconductor \cite{urech,miura}
create a spin accumulation in N.
In this paper, we consider a nonlocal spin-injection Hall device,
and discuss the anomalous Hall effect (AHE) in the presence of 
spin current (or charge current) flowing in N, 
taking into account {\it side jump} and {\it skew scattering}.

The basic mechanism for AHE is the spin-orbit interaction in N, 
which causes a spin-asymmetry in the scattering of conduction electrons 
by impurities; up-spin electrons are preferentially scattered 
in one direction and down-spin electrons in the opposite direction.  
Spin injection techniques makes it possible to induce AHE 
in {\it nonmagnetic} conductors.
When spin-polarized electrons are injected from F to N, 
these electrons moving in N are deflected by the spin-orbit scattering 
to induce the Hall current in the transverse direction and accumulate 
charge at the edges of N, yielding the spin-current induced spin Hall effect
(SHE)    \cite{hirsch,zhang,takahashiPRL88}.

%%%%%%%%%%%%%%%%%%%%%%%%%%%%%%
%\section{Formulation}\label{}
%%%%%%%%%%%%%%%%%%%%%%%%%%%%%%

Using the Boltzmann transport equations which incorporates the
spin-asymmetric scattering of conduction electrons by nonmagnetic 
impurities in N within the Born approximation, 
we can derive the ``total" spin and charge currents flowing in N
 \cite{takahashiPRL88,chapter8}
  %---------------------------------------------
  \begin{eqnarray}				
   {\bf J}_s  = {\bf j}_s + {\bf j}_s^{\rm H}, 
   \ \ \ \ \		
   {\bf J}_q  = {\bf j}_q + {\bf j}_q^{\rm H}.   
   \label{eq:Jq-Js}			
  \end{eqnarray}				
  %---------------------------------------------
where ${\bf j}_s=- ({\sN}/{e}) \ddr \dmuN$ and 
${\bf j}_q= \sN {\bf E}$ are the {\it longitudinal} spin 
and Ohmic currents, $\sN=2e^2N(0)D$ is the electrical conductivity,
$\dmuN=\frac{1}{2}(\muN^\us- \muN^\ds)$ is the chemical potential shift,
$\muN^\s$ is the chemical potential of electrons with spin $\s$,
 and $D$ is the diffusion constant.
The second terms in Eq.~(\ref{eq:Jq-Js}) are the transverse 
spin and charge Hall currents caused by spin-orbit scattering:  
  %---------------------------------------------
  \begin{eqnarray}				
  {\bf j}_s^{\rm H} &=& \aH  \left[ \hat{\bf z} \times {\bf j}_q \right]
     = \aH \sN {\,} \left( \hat{\bf z} \times {\bf E} \right),
   \label{eq:jSH} \\
  {\bf j}_q^{\rm H} &=& \aH  \left[ \hat{\bf z}\times{\bf j}_s \right]
       = - \frac{\aH\sN}{e}{\,}\left( \hat{\bf z} \times \nabla \dmuN \right), 
   \label{eq:jQH}  
  \end{eqnarray}				
  %---------------------------------------------
with $\aH=\aSJ+\aSS$, where $\aSJ = {\hbar \bar\lso}/({3mD})$ is 
the side jump (SJ) contribution, and $\aSS=({2\pi/3}){\bar\lso}N(0)\Vimp$
is the skew scattering (SS) contribution,  ${\bar\lso}=\kF^2\lso$ is the 
dimensionless spin-orbit coupling parameter, $\kF$ is the Fermi momentum,
and $\Vimp$ is the impurity potential.

Equations (\ref{eq:jSH}) and  (\ref{eq:jQH}) indicate that the spin current 
${\bf j}_s$ induces the transverse {\it charge} current (charge Hall current) 
${\bf j}_q^{\rm H}$, whereas the charge current ${\bf j}_q$ induces 
the transverse {\it spin} current (spin Hall current) ${\bf j}_s^{\rm H}$.
Equation (\ref{eq:Jq-Js}) is expressed in the matrix forms
%--------------------------------------
\begin{equation}
   \left[ 
\begin{array}{c}
  J_{q,x} \\ %%({\bf J}'_q^{\rm tot})_x   \\
  J_{s,y}    %%({\bf J}'_s^{\rm tot})_y   
\end{array}
  \right]
  =
   \left[ 
\begin{array}{cc}
       \s_{xx}      &    -\s_{xy}       \cr
       \s_{xy}      &     \s_{xx}       
\end{array}
  \right]
   \left[ 
\begin{array}{c}
  E_x  \\
  -\nabla_y \dmuN/e  \\
\end{array}
  \right],
\label{eq:matrix1}
\end{equation}
%--------------------------------------
\begin{equation}
   \left[ 
\begin{array}{c}
  J_{s,x}\\ %%({\bf J}'_s^{\rm tot})_x   \\
  J_{q,y}   %%({\bf J}'_q^{\rm tot})_y
\end{array}
  \right]
=
   \left[ 
\begin{array}{cc}
       \s_{xx}     &   -\s_{xy}  \cr
       \s_{xy}     &    \s_{xx}
\end{array}
  \right]
   \left[ 
\begin{array}{c}
  -\nabla_x \dmuN/e \\
  E_y
\end{array}
  \right] ,
\label{eq:matrix2}
\end{equation}
%--------------------------------------
where $\s_{xx}=\sN$ is the longitudinal conductivity 
and $\s_{xy}$ is the Hall conductivity contributed
from SJ and SS:
$\s_{xy}
%%= \aH\sN 
=(\aSJ+\aSS)\sN
= \s_{xy}^{\rm SJ}+\s_{xy}^{\rm SS}
%%(\s_{xy}^{\rm SJ}=\aSJ\sN, \s_{xy}^{\rm SS}=\aSS\sN)
$
with
  %-------------------------------------
  \begin{eqnarray}			
   \s_{xy}^{\rm SJ} = \frac{ e^2}{\hbar} \lso n_e, \ \ \ \ \
   %% \label{eq:sxySJ} \\
   \s_{xy}^{\rm SS} = \aSJ 
           \frac{n_e}{\nimp} \[{N(0)\Vimp}\]^{-1}, 	
   \label{eq:sxySS}			
  \end{eqnarray}			
  %-------------------------------------
where $n_e$ is the carrier (electron) density and $\nimp$ is
the impurity concentration.
Note that $\s_{xy}^{\rm SJ}$ is {\it independent} of impurity 
concentration $\nimp$.

The ratio of the SJ and SS Hall contributions is
  %-------------------------------------
  \begin{eqnarray}			
   \frac{ \s_{xy}^{\rm SJ} }{ \s_{xy}^{\rm SS} }
   = 2 \frac{\nimp}{n_e} N(0)\Vimp
   = \frac{3}{4\pi} \frac{\hbar}{\eF\timp} \frac{1}{N(0)\Vimp},
   \label{eq:ratio}			
  \end{eqnarray}			
  %-------------------------------------
where $\timp$ is the momentum scattering time and 
$\eF$ is the Fermi energy.
In ordinary non-magnetic metals, the ratio is very small because
$\nimp \ll n_e$ and $N(0)\Vimp \sim 1$, so that SS gives the
dominant contribution to SHE. 
However, in very dirty metals or in low-carrier materials such as 
doped semiconductors with $\nimp \sim n_e$, the SJ conductivity 
is comparable to or even larger than the SS conductivity in 
SHE.

%=========================================================
  \begin{figure}[b]
\includegraphics[scale=0.45]{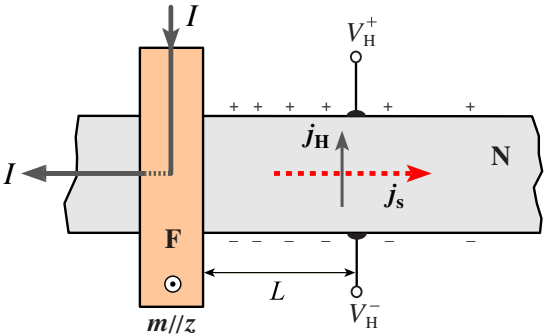}
\caption{
Spin injection Hall device (top view). The magnetic moment 
of F is aligned perpendicular to the plane.  The spin-current
induced Hall voltage $V_{\rm H}=V^+_{\rm H}-V^-_{\rm H}$ is induced 
in the transverse direction by injection of pure spin current
${\bf j}_{s}$.					
}  	\label{fig1}
  \end{figure}
%===========================================================

%%%%%%%%%%%%%%%%%%%%%%%%%%%%%%%%%%%
%\section{Nonlocal spin Hall effect}
%%%%%%%%%%%%%%%%%%%%%%%%%%%%%%%%%%%

In the following, we consider a spin-injection Hall device shown in 
Fig.~\ref{fig1}, and concentrate on the spin-current induced SHE.  
The magnetization of F electrode points to the 
$z$ direction.  When the current $I$ is sent from F to the left side of N,
the spin-polarized electrons are injected to create a pure spin current 
${\bf j}_s$ in N on the right side, 
where the total charge current is expressed as
  %-----------------------------------------------------
  \begin{eqnarray}					
   {\bf J}_{q} 
       = %%\a_{\rm H} \left[\hat{\bf z} \times {\bf j}_{s} \right]
       - ({\aH\sN}/e) \left( \hat{\bf z} \times \nabla \dmuN \right)
       + \sN {\bf E} .
   \label{eq:jtot} 				
  \end{eqnarray}					
  %-----------------------------------------------------
where the first term is the Hall current induced by 
${\bf j}_{s}$, the second term is the Ohmic current induced by surface charge, 
and $\a_{\rm H} \sim \bar{\eta}_{\rm so}N(0)V_{\rm imp}$ (skew scattering).
In the open circuit condition in the transverse direction, where
${J}^y_{q}$ vanishes, the nonlocal Hall resistance $R_{\rm H}=V_{\rm H}/I$
becomes
  %-----------------------------
  \begin{eqnarray}		
    R_{\rm H} = \frac{1}{2} \aH \PT \( \rN/\dN\) e^{-L/\lN},	
     \label{eq:VH}		
  \end{eqnarray}		
  %-----------------------------
in the case of tunnel junction, where $\PT$ is the tunneling spin polarization,
$\rN$ is the resistivity, $\lN$ is the spin-diffusion length, and
$\dN$ is the thickness of N.  In the case of metallic-contact junction
  %-----------------------------
  \begin{eqnarray}		
    R_{\rm H} = \frac{1}{2} \aH
    \frac{\pF}{1-\pF^2} 
    \( \rN/\dN\) \frac{\RF}{\RN}
    \sinh^{-1}({L/\lN}),	
     \label{eq:VH-m}		
  \end{eqnarray}		
  %-----------------------------
where $\pF$ is the spin polarization of F,
$\RN$ and $\RF$ are the spin resistances of the N and F electrodes:
$\RN = ({\rN\lN})/{\AN}$ and $\RF = ({\rF\lF})/{\AJ}$ with
$\AN$ the cross-sectional area of N and $\AJ$ the contact area between N and F.
Usually, $\RN$ is one or two orders of magnitude larger than $\RF$
\cite{takahashiPRB}.
Recently, the spin-current induced AHE have been measured
using spin injection techniques 
  \cite{kimuraJMMM,saitohAPL}.
%%\cite{kimuraJMMM,saitohAPL,valenzuela}.

%===============================================================================
\begin{table}[t]
\begin{center}
  \caption{Spin-orbit coupling parameter $\bar{\eta}_{\rm so}$ of Cu, Al, and Ag. }
  \label{table1}
{
\begin{tabular}{ccccccccc}
\hline
     & \ \ \ \ & $\lN$ (nm) & \ \ \ & $\rN\ (\mu\W$cm) & \ \ \ \ & $\timp/\tsf$  
     & \ \ \ \ \ \ & $\bar{\eta}_{\rm so}$ \\
\hline
\ Cu$^{\rm a}$  && 1000 \ \ &&  1.43  &&  0.70 $\times 10^{-3}$  && 0.040 \\
\ Cu$^{\rm b}$  && 1500 \ \ &&  1.00  &&  0.64 $\times 10^{-3}$  && 0.037 \\
\ Cu$^{\rm c}$  &&  546 \ \ &&  3.44  &&  0.41 $\times 10^{-3}$  && 0.030 \\
\ Al$^{\rm d}$  &&  650 \ \ &&  5.90  &&  0.36 $\times 10^{-4}$  && 0.009 \\
%\ Al$^{\rm e}$  &&  705 \ \ &&  5.88  &&  0.30 $\times 10^{-3}$  && 0.008 \\
\ Ag$^{\rm e}$  &&  195 \ \ &&  3.50  &&  0.50 $\times 10^{-2}$  && 0.110 \\
\hline
\end{tabular}
}
\medskip
\centerline{\small
    $^a$Ref.{\,}\cite{jedema},
    $^b$Ref.{\,}\cite{kimura},
    $^c$Ref.{\,}\cite{garzonPRL},
    $^d$Ref.{\,}\cite{jedema},
%   $^e$Ref.{\,}\cite{valenzuela}.
    $^e$Ref.{\,}\cite{godfrey}.}
\end{center}
\end{table}
%===============================================================================

It is worthwhile to make the product $\rN \lN$, which is related to
the spin-orbit coupling parameter $\bar{\eta}_{\rm so}$ as
  %---------------------------------------------
  \begin{eqnarray}                          	
    \rN \lN = \frac{ \sqrt{3}\pi}{2} \frac{R_{\rm K}}{\kF^2} 
                \sqrt{\frac{\tsf}{\timp} }
            = \frac{3\sqrt{3}\pi}{4} \frac{R_{\rm K}}{\kF^2} 
              \frac{1}{\bar{\eta}_{\rm so} } ,
     \label{eq:lso-rholso} 	               
  \end{eqnarray}                           	
  %---------------------------------------------
where $R_{\rm K}=h/e^2 \sim 25.8{\,}$k$\W$ and
$\tsf$ is the spin-flip scattering time.   The formula 
(\ref{eq:lso-rholso}) provides a method for extracting the physical 
parameters of spin-orbit scattering in nonmagnetic metals.  Using 
experimental data of $\rN$ and $\lN$ in Eq.~(\ref{eq:lso-rholso}), 
we obtain the value of the spin-orbit coupling parameter 
$\bar{\eta}_{\rm so} = 0.01$--0.04 in Cu, Al, and Ag as listed in Table~1.  
Therefore, Eqs.~(\ref{eq:VH}) and (\ref{eq:VH-m}) yields $R_{\rm H}$ 
of the order of 0.1--$1${\,}m$\W$, indicating that the spin-current 
induced SHE is observable by using nonlocal spin-injection Hall devices.

This work was supported by the NAREGI Nanoscience Project, Japan.

%%%%%%%%%%%%%%%%%%%%%%%%%%%

\end{document}